# Optimising reactive disease management using spatially explicit models at the landscape scale

Frédéric Fabre[1], Jérôme Coville[2], Nik J. Cunniffe[3]

1. UMR 1065 SAVE, INRA, Bordeaux Sciences Agro, Villenave d'Ornon F-33882, France. frederic.fabre@inra.fr
2. UR BIOSP, INRA, Domaine Saint Paul, 84914 Avignon Cedex 9, France. jerome.coville@inra.fr
3. Corresponding author. Department of Plant Sciences, University of Cambridge, Downing Street, Cambridge, CB2 3EA, United Kingdom. njc1001@cam.ac.uk

## Abstract

Increasing rates of global trade and travel, as well as changing climatic patterns, have led to more frequent outbreaks of plant disease epidemics worldwide. Mathematical modelling is a key tool in predicting where and how these new threats will spread, as well as in assessing how damaging they might be. Models can also be used to inform disease management, providing a rational methodology for comparing the performance of possible control strategies against one another. For emerging epidemics, in which new pathogens or pathogen strains are actively spreading into new regions, the spatial component of spread becomes particularly important, both to make predictions and to optimise disease control. In this chapter we illustrate how the spatial spread of emerging plant diseases can be modelled at the landscape scale via spatially explicit compartmental models. Our particular focus is on the crucial role of the dispersal kernel – which parameterises the probability of pathogen spread from an infected host to susceptible hosts at any given distance – in determining outcomes of epidemics. We add disease management to our model by testing performance of a simple "one off" form of reactive disease control, in which sites within a particular distance of locations detected to contain infection are removed in a single round of disease management. We use this simplified model to show how ostensibly arcane decisions made by the modeller – most notably whether or not the underpinning disease model allows for stochasticity (i.e. randomness) – can greatly impact on disease management recommendations. Our chapter is accompanied by example code in the programming language R available via an online repository, allowing the reader to run the models we present for him/herself.

## 1. Introduction

Diseases in crop plants can significantly impact food security (Strange and Scott, 2005), as well as production costs of food (Oerke, 2006; Savary et al. 2019) and timber (Pimentel et al. 2005). Diseases in natural environments can affect a wide range of ecosystem services (Boyd et al. 2013). Understanding when and where plant disease outbreaks are likely to occur – as well as how outbreaks can be managed effectively – is therefore imperative (Cunniffe et al. 2015a; 2016).

Many plant pathogens are extremely well established, regularly causing disease in any given location with depressing predictability, at least in the absence of crop protection. However, other pathogens – or new strains of existing pathogens – are actively spreading, leading to disease-induced losses in new regions. Our focus here is such "emerging epidemics" (Almeida, 2018), since control then has a particularly strong spatial component.

A number of high-profile emerging epidemics are currently threatening crop production worldwide, including citrus canker (Gottwald et al. 2002) and huanglongbing (Gottwald, 2010) in Brazil and the United States; cassava brown streak virus (Legg et al. 2011) and maize lethal necrosis in East Africa (Mahuku et al. 2015); Race TR4 of Panama disease in Mozambique (Ordonez et al. 2015); coffee leaf rust in South America (Talhinhas et al. 2017); novel races of wheat stem rust in Africa and the Middle East (Singh et al. 2011); and olive decline (caused by *Xylella fastidiosa*) in Europe (Martelli, 2016).

Non-native invasive pathogens in forest environments can also lead to profound population, community, ecosystem and economic impacts (Stenlid et al. 2011; Roy et al. 2014). A prominent historical example is the virtual eradication of American chestnut from the Eastern United States in the early 1900s due to chestnut blight, caused by the ascomycete *Cryphonectria parasitica* (Freinkel, 1997). Another is Dutch elm disease – caused by the beetle-vectored fungal pathogen, *Ophiostoma novo-ulmi* – which decimated elm populations in the United States and large areas of Western Europe in the 1960s and 1970s (Gibbs, 1978). Contemporary examples include *Phytophthora ramorum*, the causal agent of sudden oak death in the United States (Rizzo and Garbelotto, 2005) and ramorum disease in the United Kingdom (Brasier and Webber, 2010), as well as ash dieback (caused by *Chalara fraxinea*) across almost all of Europe (Timmermann et al. 2011).

Factors implicated in the long-distance spread of plant pathogens are numerous. Many plant pathogens, particularly those fungi that cause rust and mildew diseases, are very well adapted to spread aerially over extremely long distances (Brown and Hovmøller, 2002). Long-distance dissemination linked to the water cycle and circulation of tropospheric air masses has been described for bacteria (Morris et al. 2013). Geographical ranges of pathogens – and host plants and vectors – are changing, driven in part by climate change (Bebber, 2015). Altered patterns of trade and travel have also caused rates of pathogen introduction to increase (Brasier, 2008). New strains of pathogens resistant to fungicides – or virulent on previously resistant crop varieties – can also spread widely (Fry and Goodwin, 1997). This is particularly promoted by crop monocultures relying on single resistance genes, or combinations of resistance genes. In such cases, if even one plant becomes infected, whole fields or even regions can rapidly be lost to disease (Brown, 1995).

In facing the challenge posed by emerging epidemics, mathematical modelling can play a major role, particularly in designing and optimising management strategies (e.g. Cunniffe et al. 2015b; Thompson et al. 2018; Martinetti and Soubeyrand, 2019). This is all the more important when frequent long-distance pathogen dispersal events mean that management over large spatial scales (i.e. the "landscape" scale) must be considered (Plantegenest et al. 2007; Gilligan, 2008). Modelling offers an alternative to traditional approaches based on expert opinion, and is greatly facilitated by the progress made in the last decade in computational biology. Models provide a rational basis to integrate what is known about a pathogen with what can be learnt from early patterns of spread, while allowing for that which is not very well characterised but can be plausibly inferred from expert knowledge. The utility of such a model then lies in its ability to make predictions of future spread, which in turn can inform strategies to optimise disease detection and management (Gilligan, 2008). At the same time, however, it is imperative to recognise that a model is not a magic bullet, at least until it has been properly parameterised and validated.

For emerging epidemics, questions of practical interest are often inherently spatial. Given a certain level of resource to be expended on sampling for disease, which regions should be prioritised for surveillance? How to tailor surveillance strategies to landscape features? Once the disease has been detected, where will disease spread to next? How long will that take? How should local management be done? Which regions should be prioritised for control? Is control likely to be a success? How robust are management strategies to changing landscape contexts? These questions clearly require models to include a spatial component. Providing an introduction to the most common framework by which such questions are answered – the spatially explicit compartmental model – is one of the purposes of this chapter. An explicit intent is to make code available which allows the reader to run the model(s) for him/herself.

Emerging plant diseases are most often managed reactively, i.e. sites within a particular distance of locations detected to contain infection are targeted for control. This is often host removal, particularly for high-value crops such as fruit trees, but in principle could also be chemical treatment. The rationale is to treat or remove locations that are likely to be infected without yet showing symptoms. Taking just three real-world examples, such management is currently in progress for olive quick decline in Italy (Martelli, 2016), wheat blast in Bangladesh (Callaway, 2016) and sudden oak death in Oregon, United States (Peterson et al. 2015). It was also the basis of the decade-long attempt to eradicate sharka, a disease of prunus trees in several countries worldwide (Rimbaud et al. 2015), as well as citrus canker from Florida following its first introduction in 1995 (Gottwald et al. 2002). In the case of citrus canker, the attempt to eradicate was only stopped after being judged to have failed following removal of over 10 million citrus trees, at an estimated cost of over 1 billion dollars (Irey et al. 2006).

Such high-profile failures in control have led to a high interest from mathematical modellers. There is now a very good understanding of factors promoting the success of reactive control, including in models parameterised to the spread of particular pathogens (Cunniffe et al. 2015b; Parnell et al. 2009; 2010; Thompson et al. 2016a). More recent work has also considered how the controls can be extended to include more epidemiology, for example by including a notion of the risk of infection (Hyatt-Twynam et al. 2017; Adrakey et al. 2017), or by making controls more elaborate using tools from optimal control theory (Bussell et al. 2019). Other work has shown how reactive control can be made to be successful at very large scales even for a very well established epidemic when there is a limited budget (Cunniffe et al. 2016). However, what has not yet been tested explicitly is the effect of model structure – and in particular whether the underpinning epidemic model is deterministic or stochastic – on the extent to which a mathematical model can be used to generate the types of prediction needed to inform reactive control. Doing this, as well as providing an introduction to the theory underpinning dispersal kernels and providing a reference implementation of models in R for use by the reader, is the contribution we offer in this chapter.

## 2. Overview of the theory of dispersal kernel and the current knowledge of dispersal kernels in plant pathology

### 2.1. Mathematical classification of dispersal kernels

Many questions in both theoretical and applied epidemiology are inherently spatial. For many pathogens, transmission depends on contact between susceptible and infected individuals. In the case of sessile hosts – including plants – this in turn often depends on the distances between pairs of hosts. The movement of pathogen dispersers (e.g. spores, propagules, vectors: henceforth "inoculum") can then be described by a location dispersal kernel (Nathan et al. 2012). In this context it represents the statistical distribution of the location of the inoculum after dispersal from a point source. In two-dimensional space, the dispersal kernel can be defined as the probability density $J(x, y)$ that a propagule emitted from a point source at (0, 0) is deposited in position (x, y) (which is at a distance of $r = \sqrt{(x^2 + y^2)}$). Several families of location dispersal kernels are classical in ecology (Klein et al. 2006; Nathan et al. 2012). Although it is possible to imagine a number of deviations from this ideal in the real-world, kernels as used in models are almost always isotropic, meaning that transmission probabilities decay with distance uniformly along all radial directions.

Dispersal kernels are firstly defined by their scale, which can be taken to correspond to the mean dispersal distance. Here, we consider two families of kernels. The first is the exponential-power kernel, defined in two dimensions as

$$J_{EP}(x, y) = \frac{\tau_{disp}}{2\pi\alpha^2 \Gamma(2/\tau)} \exp\left(-\left(\frac{r}{\alpha}\right)^{\tau_{disp}}\right),$$

with $\tau_{disp}, \alpha > 0$ and so scale (i.e. mean dispersal distance) $\mu_{disp} = \alpha \Gamma(3\tau)/\Gamma(2\tau)$, and where the normalizing constant follows from integration over all of two-dimensional space. The second is the (inverse) power-law, defined in two dimensions as

$$J_{PL}(x, y) = \frac{(\tau_{disp} - 2)(\tau_{disp} - 1)}{2\pi\alpha^2} \left(1 + \frac{r}{\alpha}\right)^{-\tau_{disp}},$$

with $\alpha > 0$, $\tau_{disp} > 2$ and scale $\mu_{disp} = 2\alpha/(\tau_{disp} - 3)$.

Dispersal kernels can be further defined by their shape, which informs in particular the "fatness" of their tails. This characterizes the magnitude and frequency of long-distance dispersal events, defined, for example, as the proportion of dispersal events exceeding a given distance (e.g. the value of quantile 99% of the underpinning probability distribution for distances from the source). The shapes of several kernels sharing the same mean dispersal distance $\mu_{disp} = 80$ are illustrated in Figure 1A. Clearly, kernel shapes drastically impact the relative proportion of long-distance dispersal events. Shapes of dispersal kernels also differ markedly close to the origin, and in particular exponential-power kernels with $\tau_{disp} < 1$ are typically very strongly peaked.

The "fatness" of the kernel tail can be used to categorize kernels in a binary fashion (Mollison, 1977). When, at relatively large distance, the shape of the tail decreases less slowly than exponential distribution, or equally slowly, a kernel is termed "short-tailed" or "thin-tailed" (Klein et al. 2006). This is the case for the exponential-power kernels whenever $\tau_{disp} > 1$. Certain thin-tailed variants of the exponential-power kernel are very well known in their own right, being sufficiently well used to merit a specific name, including the Gaussian ($\tau_{disp} = 2$) and the exponential kernels ($\tau_{disp} = 1$). We note the latter kernel actually defines the frontier between thin- and fat-tailed kernels.

In contrast, if the probability of dispersal decreases more slowly than an exponential distribution at long distances from the source, kernels are termed "long-tailed" or "fat-tailed". Long-distance dispersal events are then more frequent than with an exponential kernel that shares the same mean dispersal distance. This is the case for both the exponential-power kernels with $\tau_{disp} < 1$ and for all power-law kernels. Fat-tailed kernels can be further distinguished depending on whether they are "regularly varying" (e.g. power law kernels) or "rapidly varying" (e.g. exponential-power kernels) (Klein et al. 2006). Mathematically, it implies that power-law kernels decrease even more slowly than any exponential-power function. Biologically this means that fat-tailed exponential-power kernels display rarer long-distance dispersal events than power-law kernels. As we shall see below, this distinction potentially has important implications for disease control.

*Table 1: Review of dispersal kernels obtained for plant pathogens.*

[a] The BWME kernels provide close approximations to exponential-power and power-law kernels for a wide range of parameters tested by Pleydell et al. 2018 while making the method used for parameter estimation easier. [b] Two estimates are provided, with and without considering the tail of the kernel at distances higher the radius of the experimental design. [c] The Cauchy kernel used by Neri et al. 2014 is closely related to the power-law kernels we consider.

| | Size of the design/site | Kernels considered | Anisotropy | Mean/Median (in meters) | Tail fatness | Reference |
|---|---|---|---|---|---|---|
| Sharka, Plum Pox Virus transmitted by *Aphis gosypii* | 5.6 x 4.8 km (553 orchards) | beta-weighted mixture of exponentials (BWME) kernel [a] | No | Median [CI 95%] 92.8 [82.6-104] | Fat-tailed kernel | Pleydell et al. 2018 |
| Black Sigatoka, ***Mycosphaerella fijiensis*** / Ascospores | Trap plant network of 1 km radius | Exponential-power | Yes | Mean [CI95%] 213 [144-542] D1000 14700 [2134-184267] Dinf [b] | Fat-tailed kernel | Rieux et al. 2014 |
| Black Sigatoka, ***Mycosphaerella fijiensis*** / Conidia | Trap plant network of 25 m radius | Exponential-power | Yes | Mean [CI95%] 3.15 [1.01-6.78] D25 [b] 6.12 [2.79-8.16] Dinf [b] | Thin-tailed kernel | Rieux et al. 2014 |
| Powdery mildew, ***Podosphaera plantaginis*** | 50 x 70 km (4000 meadows) | Exponential | Yes | Mean [CI95%] 860 m [640-1180] | Thin-tailed kernel (but a single kernel tested) | Soubeyrand et al. 2009 |
| Phoma stem canker, ***Leptosphaeria maculans*** Ascospores | 2.5 x 2 km | Exponential-power | No | Mean [CI95%] (years) 490 [172-1063 (2009-2010) 8.4 [5.2-12.9] (2010-2011) | Fat-tailed kernel | Bousset et al. 2015 |
| Bahia bark scaling of citrus, little is known of a putative pathogen | 420 x 212 m | Exponential kernel | No | Median [CI95%] 5 [3.92-6.42] | Thin-tailed kernel (but a single kernel tested) | Cunniffe et al. 2014 |
| Chalara ash dieback, ***Hymenoscyphus fraxineus*** | Local: traps from 0 to 800 m Regional: 15 transects from 40 to 140 km | Inverse power law, Gaussian | Yes (at regional scale) | Mean [CI95%] 1380 [640, 3320] "Local scale" 2560 [80-7650]"Regional scale" | Fat-tailed kernel | Grosdidier et al. 2018 |
| Huanglongbing, bacteria (***Candidatus*** Liberibacter) transmitted by psyllids | 3.5 x 2.4 km | Exponential, Power-law | No | Mean [CI95%, based on sd=1.5] 5 for 5 years-old trees [2-8] 10 for 18 years-old trees | Thin-tailed kernel | Parry et al. 2014 |
| Citrus canker, ***Xanthomonas axonopodis*** | 4 sites from 1 to 4 km² | Exponential, Cauchy [c] | No | Mode 50 to 120 | No significant difference between exponential and Cauchy kernel when primary infection is accounted for | Neri et al. 2014 |

## 2.2 Current knowledge of the dispersal kernel in plant pathology

Characterising dispersal kernels of plant pathogens can be very challenging. The most obvious methods require observations of inoculum dispersal patterns over large-scale ranges of distances (Kuparinen et al. 2007). Dispersal kernels can also be inferred by fitting epidemic models to disease spread data, but this requires detailed spatially explicit disease data, again over wide spatial scales and often finely time-resolved, as well as sophisticated statistical analysis (Soubeyrand et al. 2009). Consequently, relatively few studies report fitted parameters for dispersal kernels of particular plant pathogens.

Our (non-systematic) literature review identified only eight studies reporting dispersal kernels for plant pathogens that used data gathered in experimental designs extending over regions in excess of 1 km². These mostly concerned fungal pathogens, but a few focused on diseases caused by viruses and bacteria (Table 1, Figure 1B). Mean dispersal distances ranged from a few meters up to 1 kilometer, with several estimates around 100 m. However, the forms of the dispersal kernel varied, making direct comparison difficult. A study by Filipe *et al.* (2012) on sudden oak death in California which was conducted on a large area was not retained in what follows, since a full description of the fitted dispersal kernel was not given – details of short-distance (i.e. within 125 m) dispersal not being required by their underlying cell-based model.

To provide a more intuitive basis for comparison, as well as to highlight the importance of the landscape scale when addressing both basic and applied epidemiological questions, we used an individual-based approach to simulate the dispersion of propagules from a focal square field. The question is related to in-depth studies of pollen dispersion in agricultural landscapes (Lavigne et al. 2008). In all cases for computational simplicity we used the exponential kernel. Its single parameter defines (in particular) the mean dispersal distance. Assuming that infectious propagules are emitted from plants randomly scattered inside the focal infected field, we assessed, for increasing mean dispersal distances, the field area for which 50% of the pathogen propagules produced would land outside the field considered. These values are reported on the secondary x-axis of Figure 1B. A mean dispersal distance of 100 m implies that 50% of the pathogen propagules produced will land outside a field of a surface of 4.2 hectares (side length 205 m). This is higher than 2.7 ha, the median field size cultivated in France for the main cash crops (excluding market gardening, arboriculture, viticulture) (Barbu, pers. Comm., data for 2014).

Seven of the eight studies we identified used model selection approaches to distinguish between thin-tailed and fat-tailed kernels. Four of these seven studies lent support to fat-tailed kernels, including plant pathogens as diverse as viruses, fungi and oomycetes. Aerially dispersed pathogens involving, for example, spore propagules or an insect vector such as aphids escaping from plant canopy into turbulent air layer can result in long-distance flights (Ferrandino, 1993). In two of these four studies (Rieux et al. 2014; Bousset et al. 2015), exponential-power kernels were fitted to the data. Evidence for fat-tailed kernels was then derived from the confidence interval on the shape parameter ($\tau_{disp} < 1$).

These kernels are rapidly varying. Grosdidier et al. (2018) compared a short-tailed kernel to a power-law kernel, the latter being better supported by the data. These very fat-tailed kernels (with regularly varying properties) were significantly better fits to the data. Similar evidence was provided by Gibson and Austin (1996) at the field scale for an epidemic of citrus tristeza virus. This pioneering study demonstrated that a power-law relationship between infective pressure and distance is superior to an exponential one. The Gibson and Austin (1996) study also introduced to plant disease epidemiology – for the first time – the technique of data augmentation, which here facilitates writing a likelihood function and so in turn Bayesian estimation. In particular, the – unobserved – timings of infection are simply treated as additional unknown parameters to be estimated (van Dyk and Meng, 2001). However, the particular values of these parameters are then ignored via marginalisation to obtain posterior distributions of the parameters of interest. Three studies used spatially anisotropic kernels in which propagules can disperse differently depending on the direction (Soubeyrand et al. 2009; Rieux et al. 2014; Grosdidier et al. 2014). These kernels can account for the impact of local wind conditions on aerially dispersed pathogens, and so in principle allow the dispersal tail to be better captured (Savage et al. 2011).

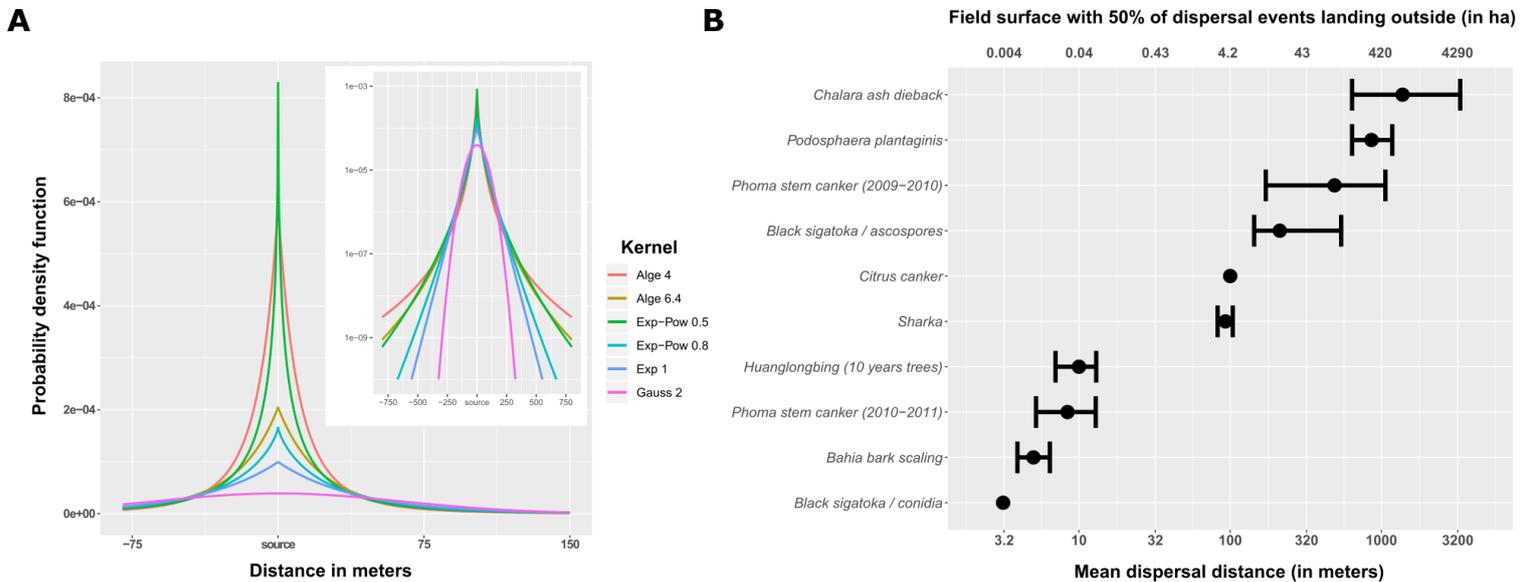

*Figure 1: Current knowledge of dispersal kernels in plant pathology.* **A:** Six dispersal kernels obtained within the power-law and the exponential-power distribution families. All have the same mean dispersal distance (here $\mu_{disp}$ = 80 meters) and differ only by the derived parameter $\tau_{disp}$ that controls the weight of the tail. In particular we show a thin-tailed Gaussian dispersal kernel ($\tau_{disp}$ = 2, quantile 99% = 194 meters), an exponential kernel ($\tau_{disp}$ =1, quantile 99% = 265 meters), two fat-tailed exponential–power kernels ($\tau_{disp}$ = 0.8, quantile 99% = 300 meters ; $\tau_{disp}$ = 0.5, quantile 99% = 404 meters) and two very-fat-tailed power-law kernels ($\tau_{disp}$ = 4, quantile 99% = 639 meters ; $\tau_{disp}$ = 6.4 , quantile 99% = 404 meters). **B:** Graphical summary of our literature review which uncovered the mean dispersal distance for eight plant pathogens. For each pathosystem, the point displays the mean (or median) and the horizontal line the extent of a 90% confidence interval (i.e. the range between the 5% and 95% quantiles of the full distribution of outputs). The secondary x-axis along the top of the figure displays, for each mean dispersal distance on the principal x-axis, and assuming that infectious propagules are emitted according an exponential kernel from plants randomly scattered inside the field, the field area (in hectares) for which 50% of the propagules produced land outside field bounds (see text). References to the particular studies used are summarised in Table 1.

# 3. The effect of dispersal kernels on epidemic dynamics and reactive host-removal-based strategies

## 3.1. Model overview

We used a simple landscape-scale epidemic model to illustrate how dispersal and model structure can affect epidemic dynamics, as well as to show how performance of a simple reactive control can be assessed. As described in the Appendix, code has been deposited in a freely available repository, allowing the reader of this article to examine the implementation and performance of the models for him/herself. Full technical details of the model are given in the Appendix; we concentrate here only upon elements required to understand the results we present.

*Table 2: Notation, model parameters, state variables and reference values.*

| Notation | Description | Ref value | Unit |
|---|---|---|---|
| Landscape description | | | |
| $n_f$ | Total number of fields | 1024 | |
| $l_f$ | Side length of an individual square field | 100 | meters |
| $p_h$ | Proportion of host fields | 0.5 | unitless |
| $A_i$ | Area of field $i$ | $10^4$ | meters² |
| Epidemic model: state variables | | | |
| $H_i(t)$ | Density of healthy plant tissue in field $i$ | na | HTD |
| $L_i(t)$ | Density of latent plant tissue in field $i$ | na | HTD |
| $I_i(t)$ | Density of infected plant tissue in field $i$ | na | HTD |
| $R_i(t)$ | Density of removed plant tissue in field $i$ | na | HTD |
| $P_i(t)$ | Plant tissue in any states in field $i$ | na | HTD |
| Epidemic model: parameter | | | |
| $m_{ij}$ | Dispersal rate from field $i$ to field $j$ | Calculated from other parameters | unitless |
| $\mu_{disp}$ | Mean dispersal distance | 80 | meters |
| $\tau_{disp}$ | Weight of the dispersal kernel tail | Explored | unitless |
| $r_h$ | Growth rate of healthy tissue | 0.2 | HTD TU$^{-1}$ |
| $K$ | Carrying capacity of each field | $10^4$ | HTD |
| $e$ | Infection efficiency | $10^{-4}$ | unitless |
| $r_p$ | Infectious propagule production rate | 2.5 | SD HTD$^{-1}$ TU$^{-1}$ |
| $\omega_L$ | Mean duration of the latent period | 7 | TU |
| $\omega_I$ | Mean duration of the infectious period | 7 | TU |
| $T_{end}$ | Duration of crop-growing | 160 | TU |
| Control strategies | | | |
| $t_{delay}$ | Time delay before replanting a field | Explored | TU |
| $r_{ctrl}$ | Radius of field removal | Explored | meters |
| $th_{ctrl}$ | Infection threshold for disease first detection | 0.2 | unitless |

TU: time unit (days); HTD: host tissue density; SD: spore density; Explored: no reference values are provided for the parameters, since the values used depend on the numerical simulation presented.

Our metapopulation model tracks pathogen spread across a landscape of discrete, square patches, each of which corresponds to a crop field. In each field we model the numbers of hosts in each of a set of epidemiological compartments, based on infection status. Patches are interconnected through a dispersal kernel, thereby parameterizing the spatial spread of the pathogen. The notation, parameters and state variables of the model are summarized in Table 2 as well as the reference values used to run simulations (unless stated otherwise in the text). The framework is inspired by Papaïx et al. (2014).

The total extent of the landscape is 3.2 x 3.2 km and it is partitioned into 1024 square fields each of area 1 hectare. We assume two distinct plant species (or varieties) are cultivated, only one of which is a host for the pathogen (Figure 2A). A proportion $p_h$ =0.5 of "host" fields contain the plant species which can be infected by the pathogen of interest; these fields are distributed at random in the landscape. The same fixed landscape is used in all the simulations presented.

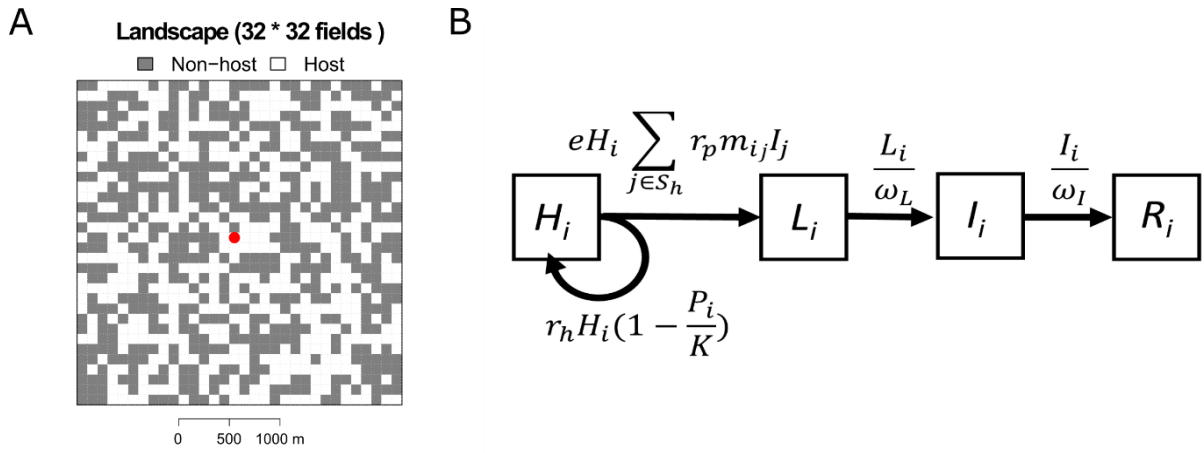

*Figure 2: Compartmental model, landscape and dispersal kernel.* **A**: Landscape of 3.2 x 3.2 km = 10.24 km² with 1024 identical square fields of 1 ha each (100 x 100 m). Host fields are shown in white; non-host fields in grey. In the accompanying code illustrating the ideas of this chapter, the landscape composition ($p_h$, proportion of host fields) and landscape composition ($a_h$, level of aggregation of host fields) can be set flexibly and independently, although in all results presented here we fix $p_h$ = 0.5 and $a_h$ = 0. All epidemics are initiated in the field marked by the red point by introducing 10 latently infected plants at *t* = 0. **B**: Schematic of the compartmental model used to describe epidemics. In each host field $i \in S_h$, host plants move from healthy (*H*) to latently infected (*L*) when first infected; from *L* to infected (*I*) after a latent period; and from *I* to removed (*R*) once the infectious period is over. *P* denotes the total density of host tissue. There is also logistic increase of healthy tissue, although in the simulations presented here this only applies following control, since the fields in our model are assumed to start at their carrying capacity of plants.

In each field, the epidemiological status of plants is represented using HLIR compartments (Figure 2B). Hosts are therefore distinguished into (H)ealthy, (L)atent (i.e. infected, but not yet able to transmit infection), (I)nfectious (i.e. infected, and able to transmit infection to other fields) and (R)emoved. Transitions between states are modelled using a system of ordinary differential equations (ODEs). The dispersal kernel allows us to estimate the net dispersal probability $m_{ij}$ of pathogen propagules between all pairs of fields (*i*,*j*) in the landscape, by integrating over all possible source and recipient host plants in both fields. Power-law and and exponential-power kernels parametrized by $\mu_{disp}$ and $\tau_{disp}$ were considered (see section 2.1).

The epidemic is initiated at *t* = 0 by introducing 0.1% of latently infected plants in a single host field located near the center of the landscape (Figure 2A). For simplicity, the same field is initially infected in all simulations. Following first infection, plant tissue remains uninfectious for a mean latent period of $\omega_L$ time units, but then becomes infectious and produces pathogen propagules for a mean infection period of $\omega_I$. These propagules can infect healthy tissue in the same field (within which the pathogen population is supposed to be perfectly mixed) but also healthy tissue in any other fields of the landscape. in all cases infection between fields *i* and *j* is weighted by the dispersal rate $m_{ij}$ as described above.

The default version of our model is based on ODEs, and so is deterministic. However, a stochastic version can readily be derived by replacing each possible transition between states in the ODE system with a single event, which occurs stochastically at a rate controlled by the corresponding term in the differential equation (Table 3; Appendix). For computational ease, binomial/Poisson draws are used to approximate the number of transitions between compartments occurring during small time intervals. We therefore use a discrete-time approximation to the underlying continuous-time model in the stochastic formulation. Examining the behaviour of this stochastic version of the model relative to the deterministic formulation allows us to systematically understand whether – and how – using a deterministic *vs.* stochastic model affects pathogen dynamics and control.

*Table 3: Transitions and probabilities for the stochastic epidemic model.* The total number of transitions of each type (with rate λ) occurring during a small time interval $(t, t + \tau)$ is drawn (i) from a binomial distribution of population size *n* and probability $p = 1 - \exp(-\lambda \tau)$ for host plants leaving a given compartment to enter another one (events healthy infection, end of latency time, end of infectious time) and (ii) from a Poisson distribution of intensity $n\lambda$ for individuals entering in a given compartment from the "outside" (event healthy birth).

| Description | Transition | Population size *n* | Rate λ |
|---|---|---|---|
| Healthy birth | $H_i \to H_i + 1$ | $H_i(t)$ | $r_h(1 - P_i/K)$ |
| Healthy infection | $\begin{cases} H_i \to H_i - 1 \\ L_i \to L_i + 1 \end{cases}$ | $H_i(t)$ | $e \sum_{j \in S_h} r_P m_{ij} I_j$ |
| End of latency time | $\begin{cases} L_i \to L_i - 1 \\ I_i \to I_i + 1 \end{cases}$ | $E_i(t)$ | $1/\omega_L$ |
| End of infectious time | $\begin{cases} I_i \to I_i - 1 \\ R_i \to R_i + 1 \end{cases}$ | $I_i(t)$ | $1/\omega_I$ |

## 3.2. The effect of the tail of dispersal kernels on epidemiological dynamics in the absence of control

We first illustrate the effect of the tail of the dispersal kernel on epidemiological dynamics (Figures 3 and S1), initially restricting our attention to the deterministic model. Epidemic dynamics are simulated for the three exponential-power kernels already plotted in Figure 1A. The kernels share the same mean dispersal distance (80 m) but are characterized by increasing tail weight ($\tau_{disp} = \{0.5, 1, 2\}$).

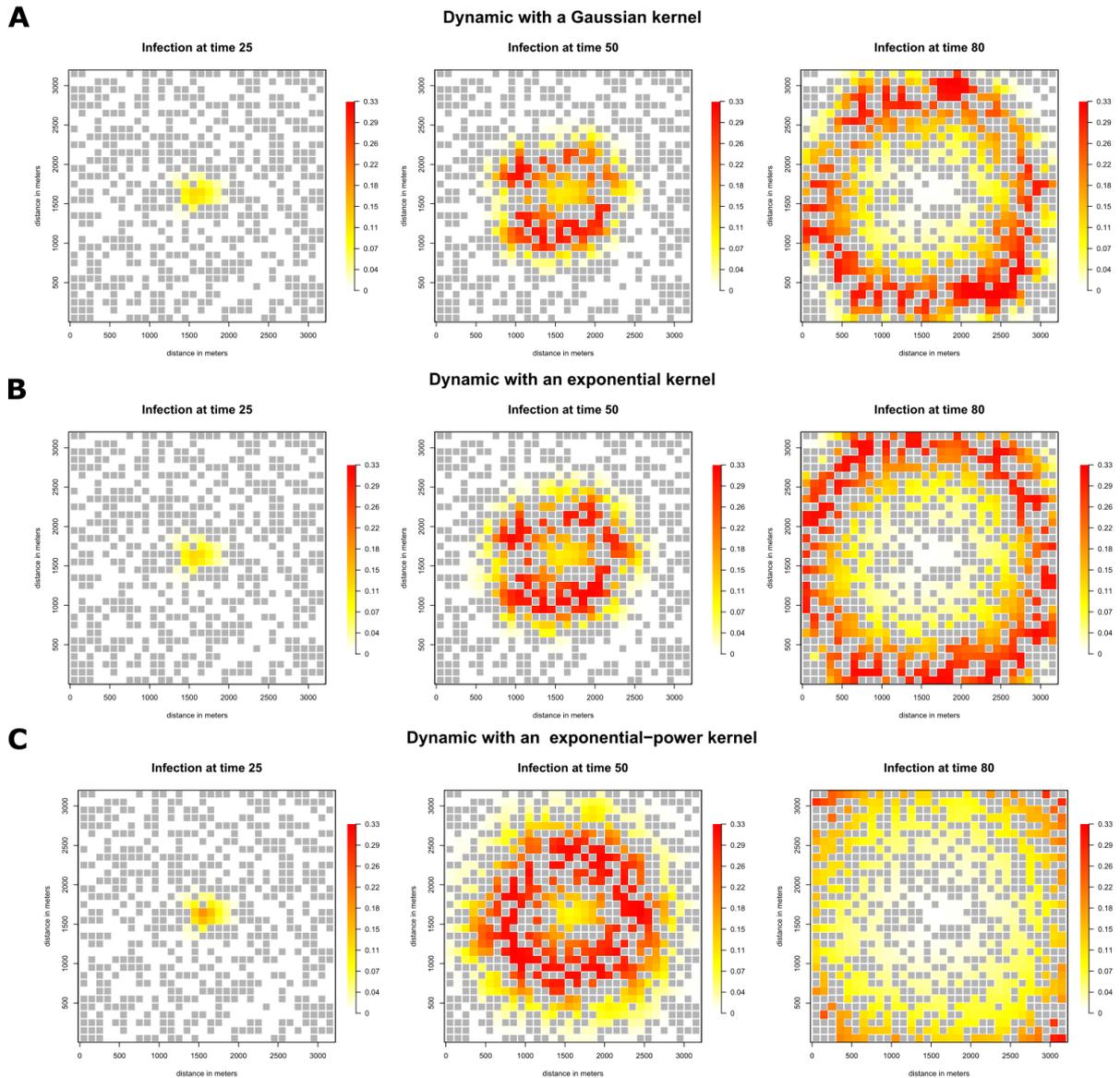

***Figure 3: Epidemic dynamics in the absence of control.*** Dynamics obtained with the deterministic model for three dispersal kernels sharing the same mean dispersal distance (80 meters) and having increasing tail weight (Gaussian, exponential and exponential-power with $\tau_{disp}$ =0.5). Epidemics are initiated in a single field near the landscape centre. Square symbols indicate non-host fields. **Line A**: Epidemic dynamics obtained with a thin-tailed Gaussian kernel ($\tau_{disp}$ =2). The proportion of infectious tissue is displayed at times 25, 50 and 80. The spreading annulus of infectious tissue surrounds a central core region in which plant tissues belong increasingly to the removed compartment (Supplementary figure S1). **Line B**: Same as line A for an exponential kernel ($\tau_{disp}$ =1).

**Line C**: Same as line A for a fat-tailed exponential-power kernel ($\tau_{disp}$ =0.5). The three kernels used are illustrated in Figure 1A.

The Gaussian and the exponential kernels both have thin tails, meaning that most disease spread occurs close to the epidemic front. An epidemic travelling wave with constant speed is then observed (van den Bosch et al. 1998) (Figure 3 A,B ; Figure S1 A,B ; Supp. video S1). When $\tau_{disp}<1$ the exponential power-kernel is not exponentially bounded. Long-distance dispersal events are therefore more frequent and induce an accelerating epidemic wave (Kot et al. 1996, Brown and Bolker, 2004;

Garnier, 2011) (Figure 3C ; Figure S1C). This type of behaviour, that has been observed for some plant diseases including wheat stem rust (caused by *Puccinia graminis* f. sp. *tritici*), southern corn leaf blight (caused by *Cochliobolus heterostrophus*) and late blight on potato (caused by *Phytophthora infestans*) (Mundt et al. 2009), is most obvious in a video of model simulations (Supp. video S1). Note that the constant speed or the accelerating epidemic waves are observed only after initial transitory dynamics within which the infection builds up locally before spreading outwards.

In the absence of control, deterministic and stochastic formulations of the model display rather similar behaviour. This is mostly because with 10 latently infected plants initially introduced (0.1% of the carrying capacity $K$ of the focal field), the probability of initial disease extinction is effectively zero. It should however be emphasized that in general the mean dynamics of the stochastic model do not exactly correspond to those obtained using a deterministic model (Allen and Allen, 2003). The major interest of the stochastic model is to allow the underlying variability of an epidemic due to demographic stochasticity to be handled (i.e. to the random variation in the number of new infection events caused by their discrete nature), as we describe below.

### 3.3 Epidemiological dynamics with host removal: is there an optimal radius?

We now focus on a disease management strategy which consists of removing plants from fields contained within a given radius around the detected epidemic focus. In particular we consider a very basic control strategy in which control is only performed a single time. Although this is extremely simple, it is the building block of the type of repeated reactive disease management that often drives both theory (e.g. Parnell et al. 2009; Cunniffe et al., 2015b; Hyatt-Twynam et al. 2017; Craig et al., 2018) and practice (Gottwald et al. 2002; Peterson et al. 2015; Martelli, 2016; Callaway, 2016). We specifically investigate how the optimal radius depends on: (i) the characteristics of the dispersal kernels of the pathogen considered and (ii) the mathematical details of the underpinning model.

We assume that disease is detectable in any field within which the sum of the proportions of infected and removed hosts exceeds 0.2. Disease control potentially occurs once per unit of time in our model (at $t$ = 1,2,3,...). At the first such time at which disease is detectable in any field across the landscape, all hosts in all fields which have centres lying within a radius $r_{ctrl}$ of any field in which disease is detectable are immediately removed. Removal is therefore initiated around one or more foci of control depending upon whether only one or multiple fields have levels of disease exceeding the detection threshold at the precise time at which control is done. As stated above, disease control occurs only once in our simplified model, and does not occur again thereafter.

We explored values of $r_{ctrl}$ ranging from 50 m, at which radius only the field(s) in which the disease was initially detected are removed, to 2000 m, at which radius the hosts in at least 98% of all fields across the landscape are removed. Removed fields were then either replanted one unit of time after removal ($t_{delay} = 1$) or were never replanted ($t_{delay} \to \infty$). This allows us to test whether and how replenishment of susceptible hosts affects the performance of disease management. We quantify performance of disease management at each control radius via the time-integrated amount of healthy tissue across the entire landscape until some notional end time $T_{end}$ as a proxy for crop yield (we present this value normalized relative to the same quantity when there is no control).

We consider first the "modeller's choice" of whether the model is deterministic or stochastic. This turns out to be of fundamental importance (Figure 4). Using the deterministic model, the disease can never be eradicated after it has been introduced. Infection spreads right across the landscape immediately from the very beginning of the epidemic, causing all fields to instantaneously contain at least some pathogen-infected host tissue, albeit sometimes at very low density. This phenomenon is called the "hair-trigger effect" in mathematics (Aronson and Weinberger, 1978), and has been referred to as the "Atto-fox" problem in epidemiology (Mollison, 1991). Consequently, whatever control radius is applied, the disease can never be eradicated. Depending on whether or not fields are replanted with healthy hosts after removal, the relative performance of control under the yield-based metric we use here therefore either increases monotonically with $r_{ctrl}$ (for $t_{delay} = 1$) (Figure 4B) or decreases monotonically with $r_{ctrl}$ (for $t_{delay} \to \infty$) (Figure 4C). The hair-trigger effect is illustrated in Figure S2. Disease is detected at time 27 and fields removed out to a radius of 800 m. At this time, a small proportion of plants is infected outside the control circle, and this allows the pathogen to continue spreading. It follows that no optimal control radius can easily be defined with the deterministic formulation of the simplified model of control we consider here. Nevertheless we note that – for particular disease-spread parameters – there might be individual values of $t_{delay}$ which allow an optimum radius to be recovered over the particular timescale of interest.

However, in the stochastic formulation of the model, epidemic eradication becomes possible (Figure 4; Figure S3). In this formulation, the epidemic spreads via discrete probabilistic disease transmission events between infected and healthy individuals. It therefore follows that the epidemic will be eradicated if no infection events have dispersed the pathogen beyond the control radius before the time of control. For a given control radius and set of disease-spread parameters, eradication is rarely guaranteed, however, although the *probability* of disease extinction increases with $r_{ctrl}$ (Figure S4). Obviously, when this probability is low (typical with power-law kernels), the stochastic and the deterministic models tend to provide fairly similar results.

With the exponential-power kernel family, including members with fat-tailed kernels, extinction becomes certain beyond a given radius in our case study (Figure 4A). The relative performance of the control strategy firstly increases with $r_{ctrl}$ and then decreases beyond a particular radius (Figures 4B, C), allowing an optimal radius of control to be defined. This is because there is a trade-off between the increased probability of disease eradication as the control radius is increased *vs.* the decreased yield that follows over-aggressively removing healthy fields. With the set of parameters used here, an optimal radius always exists irrespective of whether culled fields are replanted just after the control (Figure 4B) or are not replanted (Figure 4C). In the latter case, the highly peaked curves of yield response imply that the precise choice of the control radius is extremely important to maximize control efficiency.

The picture is very different when the two power-law kernels we considered, which we again note are very fat-tailed kernels. One percent of the dispersal events go beyond 639 m using the first power-law kernel ($\tau_{disp} = 4$) and beyond 404 m with the second kernel ($\tau_{disp} = 6.4$). The latter kernel was chosen as it shares the same quantile 99% as the exponential-power kernel with $\tau_{disp} = 0.5$. These two kernels display very similar shapes (Figure 1A). However, despite these apparent similarities, using a power-law rather than exponential-power kernel has a substantial effect on the existence of an optimal control radius, at least as we have defined optimal radius here. With the power-law kernel the probabilities of disease extinction after control are nearly zero for all radii up to 1400 m (which implies removing 61% of the host fields), and even when removing 98% of the host fields ($r_{ctrl}$ =2000 m), the probabilities of disease extinction do not exceed 0.54 ($\tau_{disp} = 4$) and 0.96 ($\tau_{disp} = 6.4$), respectively. Importantly, therefore, even in the stochastic version of our model it is effectively impossible to eradicate the pathogen in a single round of disease control with these fatter-tailed kernels. The consequent difference between the power-law and exponential kernels displaying very similar shapes is particularly striking when fields are not replanted after removal (Figure 4 C).

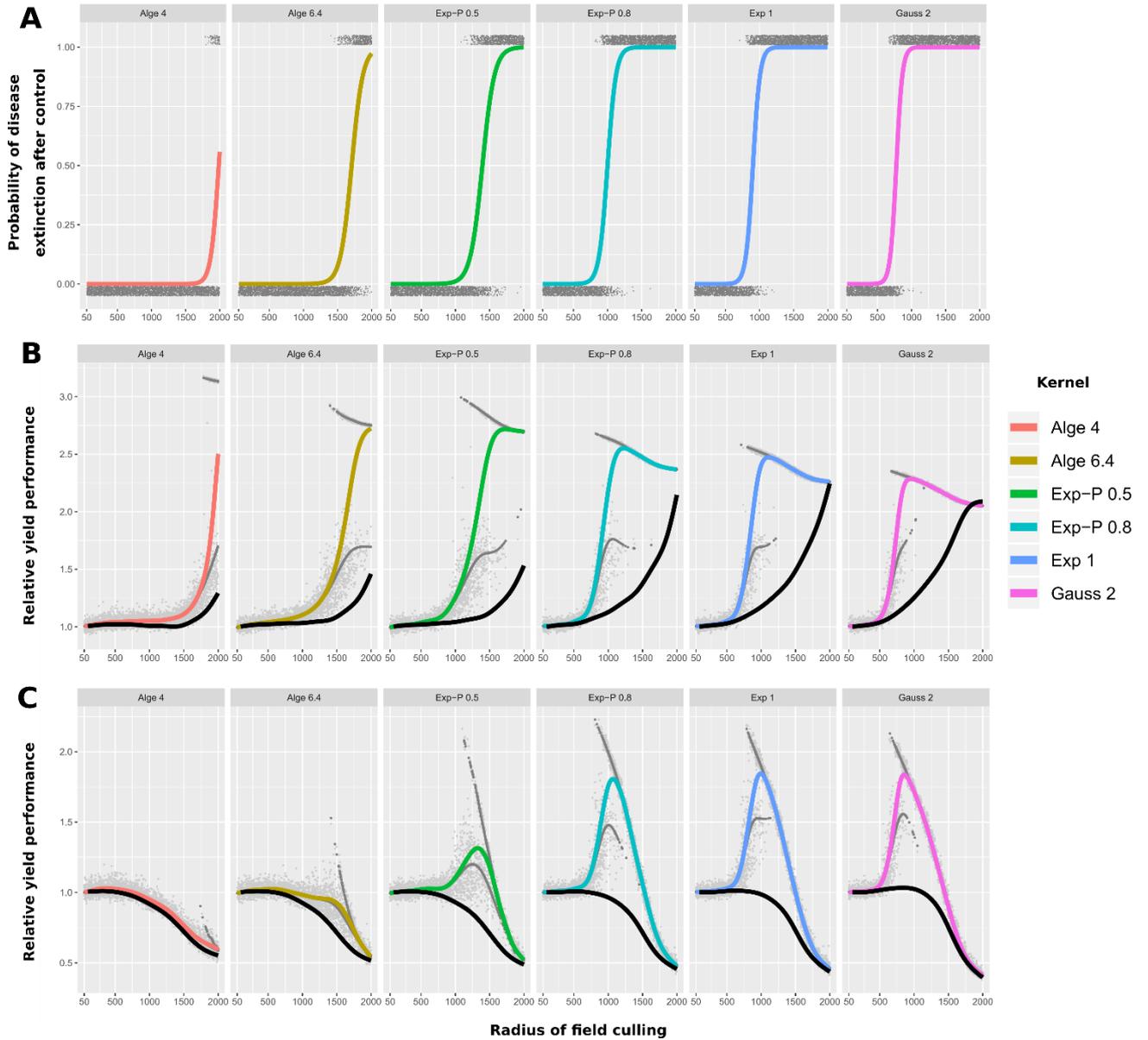

***Figure 4: Epidemic dynamics with control.*** Effects of control radius ($r_{ctrl}$), dispersal kernel and time delay before replanting ($t_{delay}$) on the probability of disease extinction after control and on the performance of control strategies. The dispersal kernels used are those plotted in Figure 1C. **Line A**: Probability of disease extinction after control (i.e. control success) as a function of the radius of field removal ($r_{ctrl}$) for the six kernels considered. **Line B**: Relative performance of each control strategy $Y^{rel}\left(r_{ctrl} \mid t_{delay}, ker\right)$ as a function of the radius of field removal for the six kernels considered with a time delay before replanting of 1 time unit $\left(t_{delay}=1\right)$. For each panel, each light grey dot corresponds to performance obtained for each individual run of the stochastic model and darker grey dots to gamma generalise additive model fitted conditionally on whether or not disease extinction occurred after control. The thick black line is the performance obtained using the deterministic model. **Line C**: Same as line B, but in the case for which there is no replanting $\left(t_{delay}=\infty\right)$.

## 4. Discussion

In this chapter, we used a simple model to highlight the influence of the interaction between pathogen dispersal kernels and modelling choice in designing optimal control strategies for an emerging epidemic. Since our ambition was simply to draw the reader's attention to a few specific points, we did not perform a full numerical exploration of our model. Nevertheless, the results we present on optimal control radii can potentially guide future research questions, as we outline below.

If initial detection of disease occurs a relatively long time after initial infection – as would be the case, for example, if disease detection was not sufficiently frequent – then the transitory phase within which infection builds up locally would be completed, and so an epidemic wave would already be spreading in the landscape. Larger control radii would then be required to achieve disease eradication, in particular with fat-tailed kernels that result in accelerating waves. However, accelerating waves do not necessarily occur when dispersal kernels are fat-tailed, but instead are conditioned upon the demographic processes involved at low densities (Alfaro and Coville, 2017). If population growth is exponential at low density (this is obviously the case with an exponential growth function but is also true with a logistic growth function), an accelerating epidemic wave is always observed with fat-tailed dispersal kernels. However, several mechanisms can prevent this type of low-density exponential growth. For example, strong Allee effects induce growth rates to become negative at low density, as a result for example of reduced fitness due to suboptimal mating opportunities (Hamelin et al. 2016). In this case, no accelerating epidemic waves can be generated from a single source, even with very fat-tailed kernels (e.g. power-law kernels). With weak Allee effects, the population growth rate always remains positive, but can become lower at lower population density. This can occur for example when the probability of infection increases with increasing parasite dose or if a threshold pathogen dose must be exceeded to overcome host basal immunity (Regoes et al. 2002). In these cases, more subtle interactions between tail fatness and per capita growth rate near zero determine if the spread is accelerating or not. In our view, studies are needed to better characterize how epidemic processes at low densities, in interaction with the form of dispersal kernels, impact reactive control.

Although here we also relied on simulations in a single landscape, using the same pathogen introduction event for each simulation, the code we present can be easily used to study how landscape features impact reactive control. Initial propagation of a newly introduced disease can be strongly impacted by local landscape features (Ostfeld et al. 2005; Plantegenest et al. 2007; Meentemeyer et al. 2012; Papaïx et al. 2014), particularly in interaction with pathogen dispersal. Basic characteristics of a landscape include its composition (i.e. the proportion of different types of habitat, including the fraction of plants that can act as pathogen hosts – defined by $p_h$ in this study) and its configuration (i.e. the specific spatial arrangement of habitat). The framework proposed can easily handle different landscape configurations resulting from varying levels of aggregation of host fields.

The third, fairly severe, restriction on the results presented here is that only a single round of reactive control was considered, i.e. in our simulations, hosts are removed only at one single time. Whenever the disease escapes eradication in this single round of control, the pathogen therefore spreads unperturbed thereafter. As we have seen, and particularly in the deterministic formulation of the model, this has profound effects on the efficacy of disease management. It even means that for our simple situation an optimal radius could not be defined in our stochastic model when there was a very fat-tailed power-law dispersal kernel, because eradication became effectively impossible. However, in practice, disease is controlled more than once, with multiple rounds of reactive removal. Even if disease is not eradicated, the amount of infection in the system can be greatly reduced by each single round of control. It is therefore possible that repeated controls could "damp down" transmission sufficiently to reduce the (effective) basic reproduction number of the epidemic to smaller than one, thereby controlling the epidemic. This would be concordant with a number of theoretical studies which show that effective reactive control of diseases is possible, even when they spread according to an extremely fat-tailed kernel. Examples of model-based studies showing that control of such epidemics can be successful include models of the animal disease, foot-and-mouth (Tildesley *et al*. 2006) and a range of plant diseases, including citrus canker (Cunniffe *et al*. 2015b), huanglongbing (Hyatt-Twynam *et al*. 2017; Craig *et al*. 2018) and sudden oak death (Filipe *et al*. 2012; Cunniffe *et al*. 2016). These types of studies also tend to highlight the importance of effective disease detection in promoting successful control (Thompson *et al*. 2016b; Parnell *et al*. 2015,2017), since effectively detecting disease means there is a smaller problem to be solved at the time of disease management (Epanchin-Niell *et al*. 2012).

More generally, we would like to draw the reader's attention to two main points. The first concerns what we have called the "modeller's choice" in deciding to use a deterministic or a stochastic formulation of the model. We have shown that the deterministic model we considered here might not be suitable to generate the types of prediction needed to inform reactive control. Generally speaking, it illustrates that deterministic formulations often become inappropriate at low population densities (Renshaw 1991), via what is sometimes referred to as the "Atto-fox" problem (Mollison, 1991; the name comes from unrealistic recovery in infected densities following near eradication in early deterministic models of the spatial spread of rabies in foxes). Predictions concerning the efficacy of spatially explicit reactive control strategies derived via deterministic models therefore require careful interpretation. However, deterministic models can be adapted to allow control to be represented. One obvious way to do so is to relax the formulation of the model to allow epidemic extinction to occur with a deterministic model, as a result of specific interaction between demographic processes and dispersal kernel properties. We note in passing that another choice faced by modellers using deterministic models is whether to use integro-differential equations (IDE), that explicitly represent pathogen dispersal using kernels, or partial differential equations (PDE), that represent pathogen dispersal by a diffusion process. For thin-tailed kernels the spreading dynamics obtained with IDE or PDE approaches do not differ at least qualitatively (Schumacher, 1980; Weinberger, 1982; Medlock and Kot, 2003; Coville et al., 2008). However, the underlying dispersal processes differ and are distinguishable statistically. In particular it is important to emphasize that continuous-time IDE models with Gaussian kernels are in essence different from reaction-diffusion models.

Our second point is that many more studies are needed to inform dispersal kernels of plant pathogens. Several aspects must be taken into consideration. First, in the set of eight studies selected based on the size of experimental design, no studies compared fat-tailed *vs.* very fat-tailed kernels, and more generally none presented results concerning more than two families of dispersal kernels when fitting data. Doing such a comparison is clearly interesting, particularly since we have seen here that the precise characterization of the fat-tail weight drastically impacts applied issues such as defining an optimal control radius. However, and this is the second point here, designing experiments to precisely infer dispersal kernels is clearly challenging given the order of magnitude of dispersal of many plant diseases. Accordingly, we only reviewed studies using data gathered in experimental designs of 1 km² or greater. It is well known that data confined to relatively small spatial scales can blur the precise estimates of the form of dispersal at large distances, and in particular the shape of the kernel's tail (Ferrandino, 1996). Indeed, the spatial scale at which observations are realized is a major concern when fitting and comparing dispersal kernels. Kuparinen et al. (2007) showed how kernels with very different tails may yield similar fits. They found that the predicted dispersal at long distances depends on both the kernel considered and the distances over which the dispersal data was collected. Moreover, observing a single realization of a dispersal process is not enough to inform the dispersal process. Rather, dispersal outcomes have to be observed under varying spatio-temporal conditions (Kuparinen et al. 2007; Nathan et al. 2012). This is firstly because the basic dispersal process varies according to both biotic factors (genetic effects, plant canopy structure, vector behaviour, etc.) and abiotic factors (landscape features, weather conditions, etc.). In this regard, the difference observed in the dispersal kernels of *Leptosphaeria maculans* for two consecutive transitions from one season to the next – probably due to differences in both climatic conditions and the forces of the inoculum sources – is striking (Bousset et al. 2015). This is also because properly estimating the uncertainty associated with parameters of kernels is an important task.

A first direction to facilitate estimation of pathogen dispersal is to use new sources of host and disease spread data. Integrating new sources of data at different scales will help to better resolve host and pathogen locations. Promising sources of data include unmanned aerial vehicles, remote sensing, and earth observation. There have been recent high-profile successes in detection of a single pathogen from aerial imagery (Zarco-Tejada et al. 2018). However, accurately distinguishing symptoms caused by different pathogens – as well as distinguishing disease from a more general signature of "stress" – is expected to remain rather challenging (Mahlein, 2016).

There are also exciting possibilities that follow from better integrating genomic data into epidemiology. Methods for parameterizing pathogen dispersal use data augmentation to integrate over all chains of transmission consistent with observed spread data (Gibson and Austin, 1996). Studies of human (Jombart et al. 2014) and animal (Ypma et al. 2012) pathogens show how genomic data can be used to constrain chains of transmission more tightly and so improve the precision of model fits. However, for plant disease models, integration with genetic information is in its very early stages (Picard et al. 2017).

A second direction to improve estimation of pathogen dispersal is to describe dispersal pathways in models in a more realistic way (Cunniffe et al., 2015a). More complex dispersal kernels could be included in forward simulations relatively easily, and are already available for some pathways. These include atmospheric dispersion models (Singh et al. 2011; Meyer et al. 2017) and spread via trade networks (Shaw and Pautasso, 2014). However, attention must be paid to understanding whether including these pathways leads to more accurate prediction, since it is possible that underpinning models will become rather complex.

Recently, Leyronas et al. (2018) provide evidence that the arrival in a given area of airborne inoculum of *Sclerotinia sclerotiorum* from remote origins can be predicted using connectivity networks of air mass movements in the troposphere. In their approach, the directional connectivity between a particular pair of source and sink sites is estimated using archived meteorological data provided by the Global Data Assimilation System (GDAS) of NOAA and the software HYSPLIT that models air-mass trajectories. They also provide evidence that directional connectivity is more informative than the simple geographic distance. This study – as well as other similar studies based on long-distance dispersal of the wheat stem rust pathogen (Meyer et al. 2017) – opens new avenues to understand the atmospheric highways of airborne pathogen dispersal and, from an applied perspective, new opportunities to set up surveillance networks (Carvajal-Yepes *et al*. 2019).

# APPENDIX

**Detailed description of our model**

The notation, parameters along with their reference values, and state variables of the model are summarized in Table 2. The R code is freely available at https://doi.org/10.15454/JWONRG and mainly relies on the package "deSolve" (for solving differential equations) and "raster" (to describe a spatially explicit agricultural landscape).

**Deterministic formulation of the model**

We perform simulations on a single host landscape for all simulations, using the same initial condition in each (Figure 2A). Let $S_h$ denote the set of indices of the fields containing host plants. Initially (at *t*=0), only healthy plants are cultivated at density $K$ in each host field ($H_i(0) = K$ for $i \in S_h$ ; $H_i(0) = 0$ elsewhere in non-host fields). We note that the host population does not necessarily represent individual plants but can also be viewed as leaf area densities (leaf surface area per m²). Although it is only relevant when fields are replanted with healthy hosts following disease in our default use of the model, in the absence of disease the density of healthy plant tissue grows logistically at rate $r_h$ with carrying capacity $K$ (which we assume is identical for each field, since in our model all fields have the same area).

The epidemic is initiated by introducing 10 (<< *K*) latently infected individuals in a host field located near the center of the landscape (Figure 2A). The epidemic runs from $t=0$ to $T_{end}$. Transitions between the four states considered [(H)ealthy, (L)atent (i.e. infected, but not yet able to transmit infection), (I)nfectious and (R)emoved] are modelled using a system of ordinary differential equations. In the following we denote $P_i = H_i + E_i + I_i + R_i$ the total density of hosts. These equations rely on parameters classically used in botanical epidemic models (infection efficiency *e*, rate of production of infectious propagules *r_p*, mean duration of the latency ($\omega_L$) and sporulation ($\omega_I$) periods) (Madden et al. 2007). We use a metapopulation model in which each patch is a single field. Each field displays homogeneous habitat conditions and, in particular, the pathogen population is supposed to be perfectly mixed within a patch. The pathogen populations are linked between host fields via dispersal. A matrix of derived parameters (with individual elements $m_{ij}$) parameterizes the net dispersal probability of pathogen propagules between each pair of fields (i,j) in the landscape. Given a dispersal kernel, the individual elements $m_{ij}$ can be estimated by integrating over all possible source and recipient plants in both fields (see following section). For landscapes or fields with more complex topologies (e.g. fields containing obstacles as for example a lake), dispersal rates are better approximated using a dedicated algorithm such as CaliFloPP (Bouvier et al. 2009). The model of the density of tissue in each epidemiological compartment in field *i* is therefore

$$\begin{cases} \dfrac{dH_i}{dt} = r_h H_i(1-\dfrac{P_i}{K}) - eH_i \sum_{j \in S_h} r_P m_{i,j} I_j \\ \dfrac{dL_i}{dt} = eH_i \sum_{j \in S_h} r_P m_{i,j} I_j - \dfrac{1}{\omega_L} L_i \\ \dfrac{dI_i}{dt} = \dfrac{1}{\omega_L} L_i - \dfrac{1}{\omega_I} I_i \\ \dfrac{dR_i}{dt} = \dfrac{1}{\omega_I} I_i \end{cases}$$

Note that for each host field, the total host density $P_i$ follows a logistic curve with carrying capacity $K$. As a consequence if $P_i(0)=0$ or $P_i(0)=K$ then for all time $t \geq 0$ $P_i(t)=0$ or $P_i(t)=K$ and the above system reduces to a simpler model without logistic growth. For such a model the equilibrium are (i) $(K,0,0,0)$ or (ii) $(0,0,0,K)$ for each host fields. Only the latter is stable for the dynamics (*i.e.* as soon as exposed or infected plants are introduced, all the compartments converge to this equilibrium).

**Estimation of the dispersal rate between pairs of fields**

Given a kernel, one can derive the dispersal rate between any pair of fields *i* and *j* $(1 \leq i \leq j \leq n_f)$ in the landscape as

$$m_{i,j} = \dfrac{1}{|A_i|} \int_{A_i} \left( \int_{A_j} J(x-x', y-y') dx' dy' \right) dx dy ,$$

where $|A_i|$ is the area of field *i*.

By assimilating all the points of the fields *i* and *j* to their centroids $(x_i, y_i)$ and $(x_j, y_j)$, the rate can be crudely approximated by

$$m_{i,j} \approx \bar{m}_{i,j} = J(x_i - x_j, y_i - y_j)|A_j|,$$

with an error

$$Error(i,j) \sim \dfrac{|A_j|}{|A_i|} \int_{A_i} \left( |x-x_i| + |y-y_i| \right) dx dy + \int_{A_j} \left( |x-x_j| + |y-y_j| \right) dx dy .$$

For identical square fields, the error is $Error(i,j) \sim l_f^3$. The estimation errors can be decreased by partitioning (i) sub-regions inside fields and (ii) computing dispersal rates between these elementary cells. For $A_i = \bigcup_{k=1}^{n_i} A_{i,k}$ and $A_j = \bigcup_{l=1}^{n_j} A_{j,l}$, we then have

$$\bar{m}_{ij} = \dfrac{1}{|A_i|} \sum_{k=1}^{n_i} \sum_{l=1}^{n_j} \bar{m}_{k,l} |A_{i,k}|,$$

where $n_i$ (resp. $n_j$) is the number of sub-regions of the field *i* (resp. *j*), $|A_{i,k}|$ is the area of the subregion $A_{i,k}$ and $\bar{m}_{k,l}$ is the approximation computed with the centre of each sub-region, i.e.

$$\bar{m}_{k,l} = J(x_{i,k} - x_{j,l}, y_{i,k} - y_{j,l}) |A_{j,l}| .$$

where $(x_{i,k}, y_{i,k})$ and $(x_{j,l}, y_{j,l})$ are the positions of the centroids of the subregions $A_{i,k}$ and $A_{j,l}$, and $|A_{j,l}|$ is the area of the subregion $A_{j,l}$.

When $n_i = n_j = n_s$ and the sub-regions are subsquares such that $|A_{i,k}| = |A_{j,l}| = (l_f/n_s)^2$ the error in the computation is then of order $(l_f/n_s)^3$.

**Discrete stochastic formulation of the model**

To obtain a stochastic formulation of the model, each transition in the differential equation is replaced by a single event, each of which occurs stochastically at a rate controlled by the term in the differential equation. The population size of each compartment takes only integer values {0,1,2,…,K}. For computational ease, time is discretized in our simulations, using independent binomial/Poisson draws to calculate the number of transitions between compartments occurring during the time interval $(t, t + \tau)$ (Table 3).

**Modelling control**

As described in the main text, only a very simple control strategy based on a single round of host removal is considered. We assume that disease can be detected in any field in which the sum of the proportions of infected and removed hosts exceeds $th_{ctrl}$. Detection and control occurs once, at time $t_{detect}$, which is constrained to be an integer value (i.e. the first integer time falling after any time at which the level of disease exceeds in any field). At this time, all fields with centres within a radius $r_{ctrl}$ of any detectable field are removed. Removed fields are sown again at time $t_{detect} + t_{delay}$ with an initial density of healthy plants of $0.01 \cdot K$. Contrary to the case without control, the logistic growth is at play in the dynamics.

**Statistical modelling of the performance of control**

For each combination of the time delay before replanting ($t_{delay}$, 2 levels) and dispersal kernel ($ker$, 6 levels), we performed $n_{sim} = 2000$ simulations of the stochastic model by sampling $n_{sim}$ control radii $r_{ctrl}$ from a continuous uniform distribution U[50,2000]. For each stochastic run, disease extinction occurred, or not, after the control. These $n_{sim}$ Bernoulli outcomes were used to estimate the probability of control success (i.e. probability of disease extinction following control) $p(r_{ctrl} | t_{delay}, ker)$ as a function of $r_{ctrl}$ using a logistic regression (R function glm, family=Binomial).

Secondly, for each stochastic run, the relative performance of a control strategy with radius $r_{ctrl}$ was assessed using the relative yield, $y^{rel}(r_{ctrl} | t_{delay}, ker) = y(r_{ctrl} | t_{delay}, ker) / y_\emptyset(ker)$, where $y(r_{ctrl} | t_{delay}, ker) = \sum_{i \in S_h} \int_0^{T_{end}} H_i(t)$ is the time-integrated density of healthy plants over all fields (a proxy of crop yield; see e.g. Jeger (2004)) with control, and $y_\emptyset(ker)$ is the same without control.

We then used generalized additive models with gamma family to estimate mean relative performances as a function of the control radius (R package mgcv, function gam, family=Gamma(link=log)). In particular, we fitted two models, one for the set of simulations within which control was "successful" (i.e. disease eradication occurred), and another for the set in which control was "not successful" (i.e. disease was not eradicated). For successful control, the proportion of deviance in the expected relative performance conditional on successful control $Y^{rel}_{sucess}\left(r_{ctrl} \mid t_{delay}, ker\right)$ explained by the 12 models (2 levels for $t_{delay}$ * 6 levels for $ker$) ranged from 0.3 to 0.99 (mean=0.92, sd=0.2). For the models conditional on control failure, $Y^{rel}_{failure}\left(r_{ctrl} \mid t_{delay}, ker\right)$, the proportion of the deviance explained by the 12 models fitted ranged from 0.51 to 0.97 (mean=0.79, sd=0.1).

Finally we estimated the mean relative performance of a control strategy with radius $r_{ctrl}$ as

$$Y^{rel}\left(r_{ctrl} \mid t_{delay}, ker\right) = Y^{rel}_{sucess}\left(r_{ctrl} \mid t_{delay}, ker\right).p(r_{ctrl} \mid t_{delay}, ker)$$
$$+ Y^{rel}_{failure}\left(r_{ctrl} \mid t_{delay}, ker\right).\left(1 - p(r_{ctrl} \mid t_{delay}, ker)\right)$$

# SUPPLEMENTARY MATERIALS

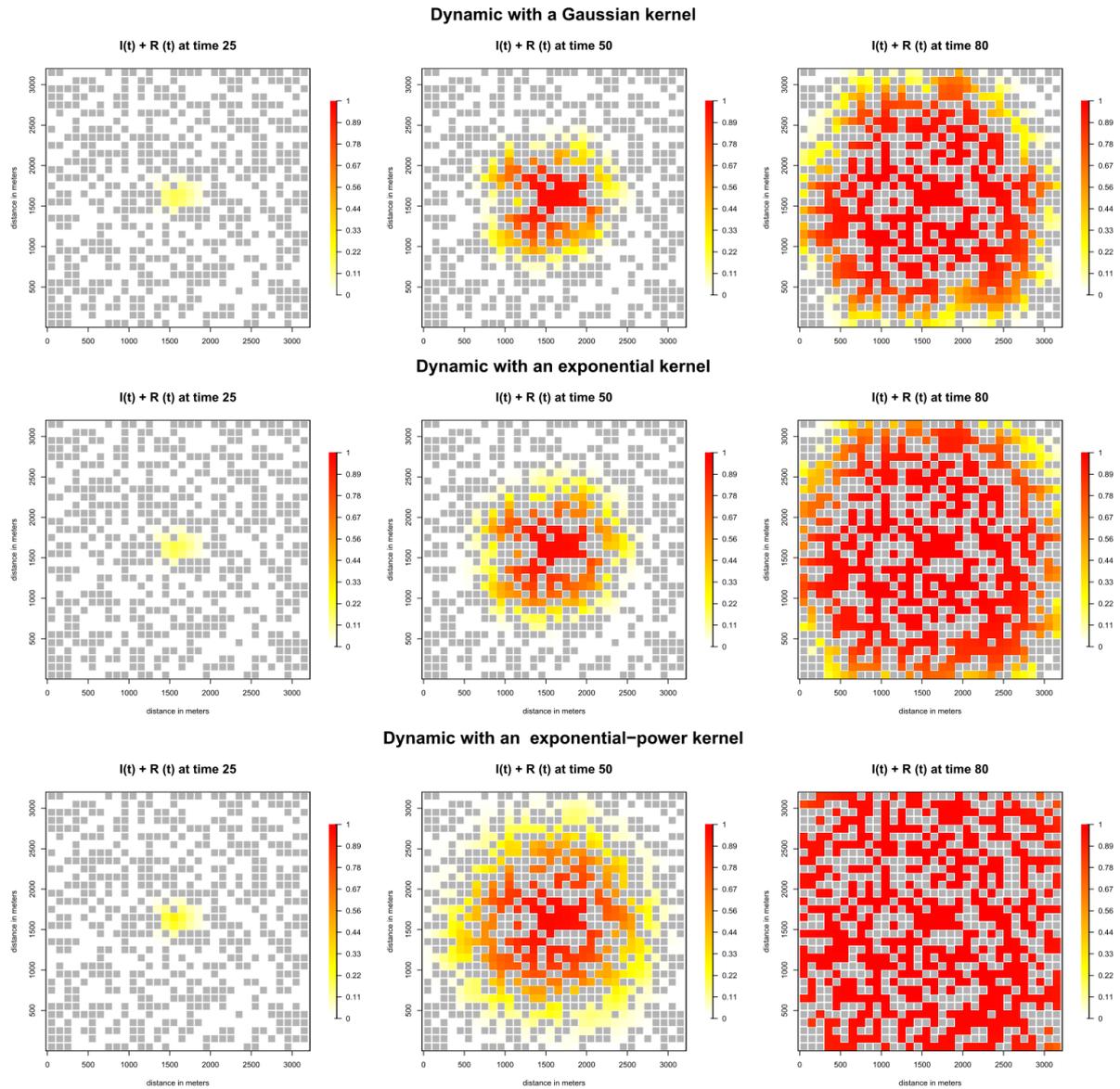

**Supplementary figure S1:** Epidemic dynamics obtained for three dispersal kernels sharing the same mean dispersal distance (80 m) and having increasing tail weight (Gaussian, exponential and exponential-power with $\tau_{disp}$ =0.5). This figure is the same as Figure 3 except the sum of the proportions of infectious and removed tissue is reported.

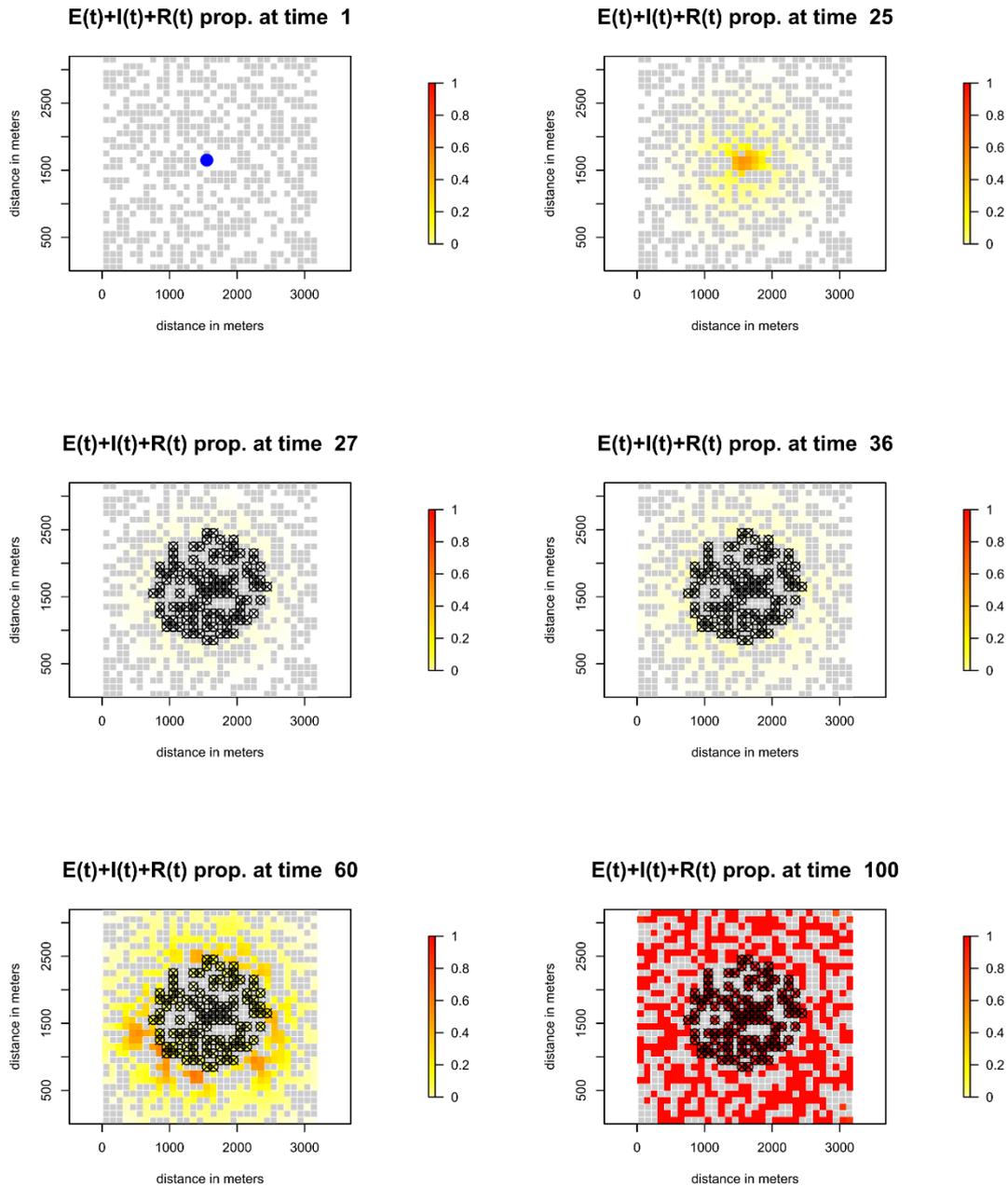

**Supplementary figure S2:** Epidemic dynamics with a control radius of 800 m obtained using the deterministic model, with an exponential kernel with a mean dispersal distance of 80 m. Control occurred at time 26. The sum of the proportions of exposed, infectious and removed plants is displayed at several times before and after control. At time 1, the blue point locates the field in which the epidemic is initiated. After control, the circle-cross points locate culled fields. Removed fields were replanted one day after control. With the deterministic model, the epidemic is not eradicated by the control.

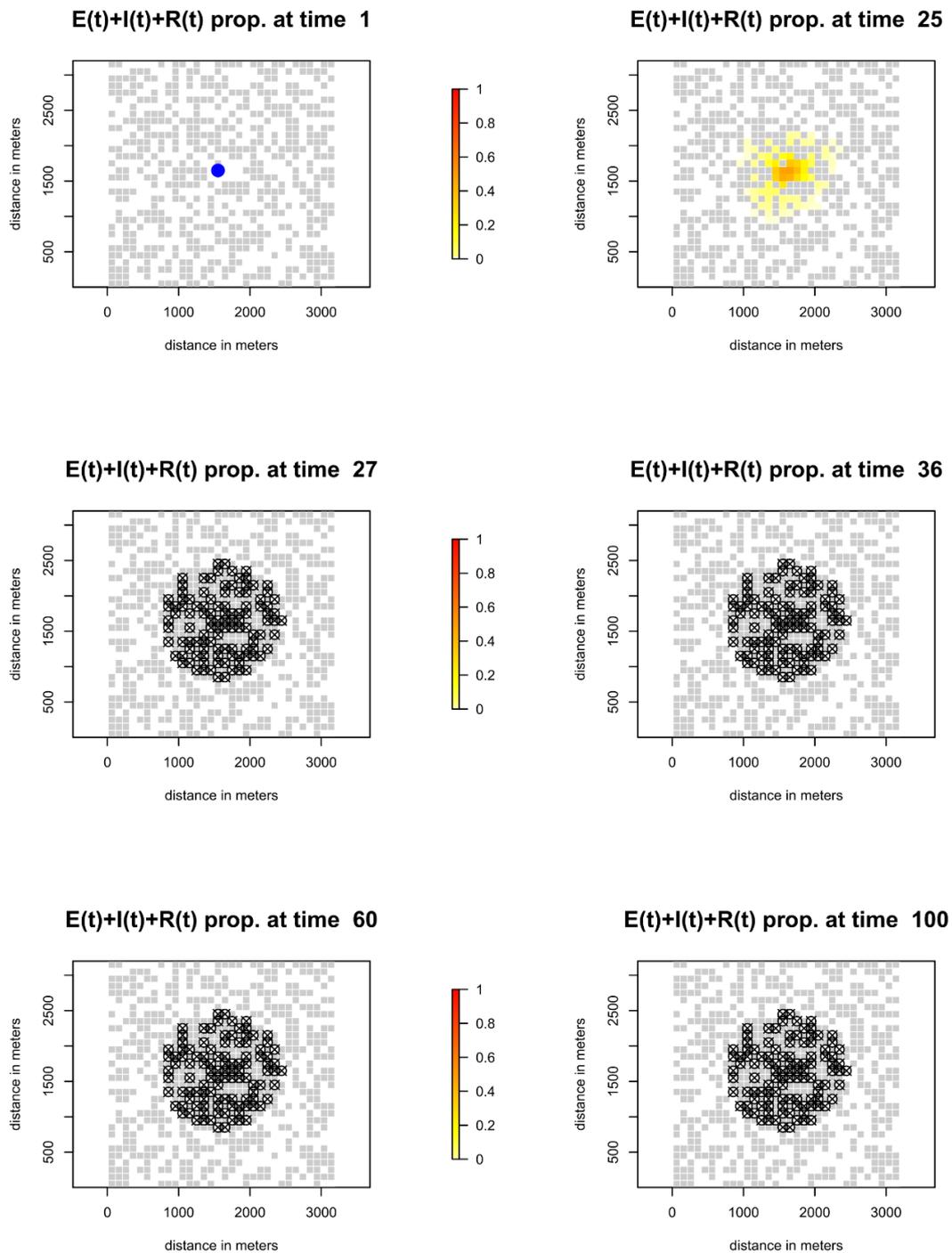

**Supplementary figure S3:** Epidemic dynamics with a control radius of 800 m obtained with one realization of the stochastic model in which the epidemic was eradicated by the control. Control occurred at time 27. The sum of the proportions of exposed, infectious and removed plants is displayed at several times before and after control. At time 1, the blue point locates the field in which the epidemic is initiated. After control, the circle-cross points locate culled fields. Removed fields were replanted one day after control. An exponential kernel with a mean dispersal distance of 80 m is again used.

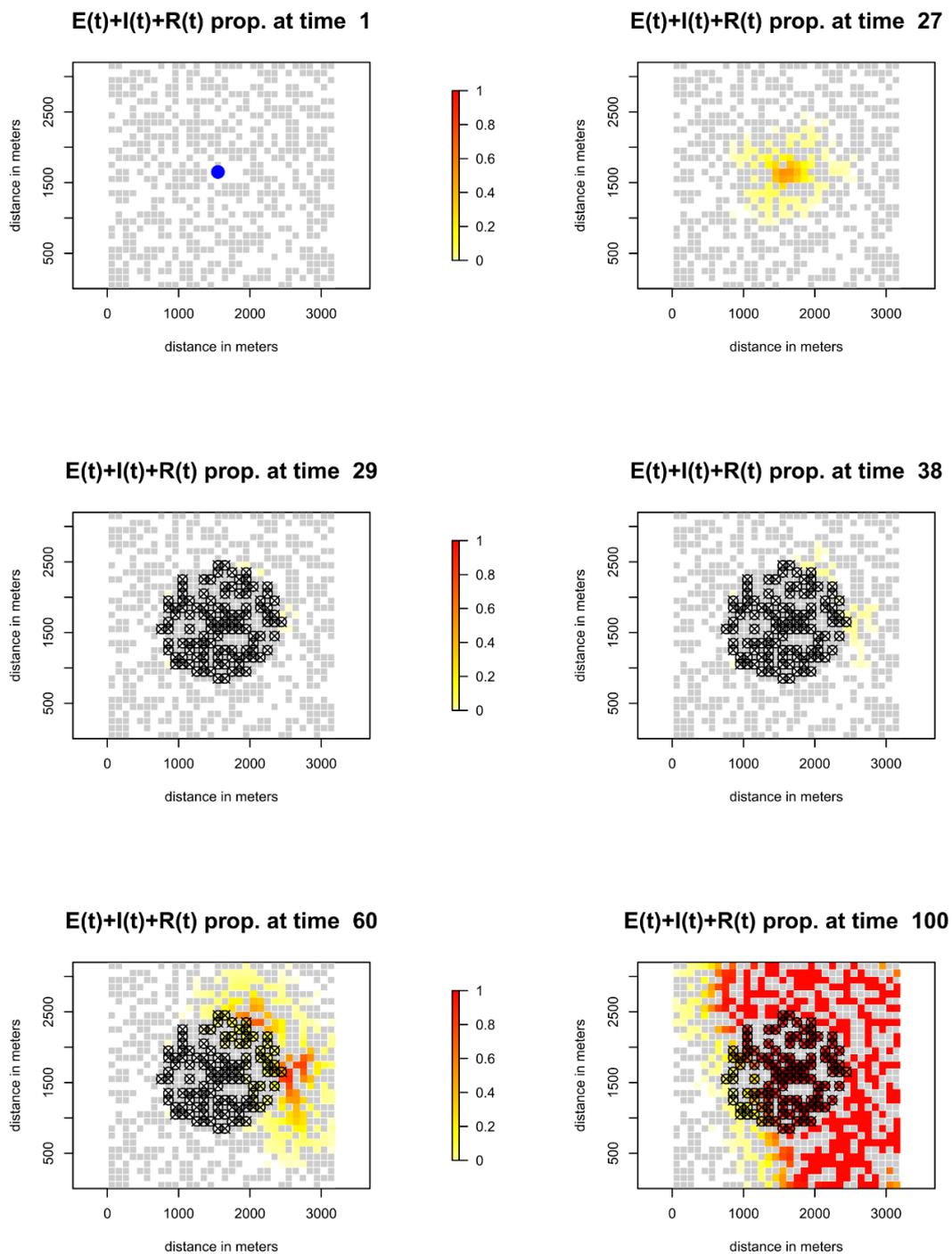

**Supplementary figure S4:** Epidemic dynamics with a control radius of 800 m obtained with one realization of the stochastic model in which the epidemic was not eradicated by the control. Control occurred at time 27. The sum of the proportions of exposed, infected and removed plants is displayed at several times before and after control. At time 1, the blue point locates the field in which the epidemic is initiated. After control, the circle-cross points locate culled fields. Removed fields were replanted one day after control. An exponential kernel with a mean dispersal distance of 80 m is again used.

[to be downloaded at https://doi.org/10.15454/JWONRG]

**Supplementary video S1:** Epidemic dynamics obtained with the deterministic model for two exponential–power dispersal kernels sharing the same mean dispersal distance (80 m) and differing by their tail weight. The first kernel has a thin tail ($\tau_{disp}$ =2, Gaussian case). The second kernel has a fat tail ($\tau_{disp}$ =0,5). The proportion of infectious tissue only is displayed in the upper line of 3 graphs while the sum of the proportions of infectious and removed tissue is displayed in the lower line. Note scales of the colours used to show proportions differ between the top and bottom lines: in the upper line, dark red corresponds to 0.33, whereas in the lower line, dark red corresponds to 1.0.